\documentclass[12pt]{spieman}  % 12pt font required by SPIE;

\usepackage{amsmath,amsfonts,amssymb}
\usepackage{graphicx}
\usepackage{setspace}
\usepackage{tocloft}

\title{Mitigation of artifacts due to isolated acoustic heterogeneities in photoacoustic computed tomography using a variable data truncation-based reconstruction method}

\author[a]{Joemini Poudel}
\author[a]{Thomas P. Matthews}
\author[a]{Lei Li}
\author[a]{Mark A. Anastasio}
\author[a]{Lihong V. Wang}
\affil[a]{Washington University in St. Louis, Department of Biomedical Engineering, 1 Brookings Dr, St. Louis, USA, 63130}

\cftpagenumbersoff{figure}
\cftpagenumbersoff{table} 
\begin{document} 
\maketitle

\begin{abstract}
			Photoacoustic computed tomography (PACT) is an emerging computed imaging modality that exploits optical contrast and ultrasonic detection principles to form images of the absorbed optical energy density within tissue.  If the object possesses spatially variant acoustic properties that are unaccounted for by the reconstruction method, the estimated image can contain distortions. While reconstruction methods have recently been developed for compensating for this effect, they generally require the object's acoustic properties to be known \textit{a priori}.  
			To circumvent the need for detailed information regarding an object's acoustic properties, we have previously proposed a half-time reconstruction method for PACT.  A half-time reconstruction method estimates the PACT image from a data set that has been temporally truncated to exclude the data components that have been strongly aberrated. However, this method can be improved upon when the approximate sizes and locations of isolated heterogeneous structures such as bones or gas pockets are known.  
			To address this, we investigate PACT reconstruction methods that are based on a variable data truncation (VDT) approach. The VDT approach represents a generalization of the half-time approach, in which the degree of temporal truncation for each measurement is determined by the distance between the corresponding ultrasonic transducer location and the nearest known bone or gas void location.  Computer-simulated and experimental data are employed to demonstrate the effectiveness of the approach in mitigating artifacts due to acoustic heterogeneities.   
\end{abstract}

% Include a list of up to six keywords after the abstract

% Include email contact information for corresponding author
{\noindent \footnotesize\textbf{*}Mark A. Anastasio,  \linkable{anastasio@wustl.edu} \newline \footnotesize\textbf{*}Lihong V. Wang,  \linkable{lhwang@wustl.edu} }

\begin{spacing}{2}   % use double spacing for rest of manuscript

\section{Introduction}
		\label{sec:intro}  % \label{} allows reference to this section
		Photoacoustic computed tomography (PACT) is an emerging imaging modality that combines the high optical contrast of blood-rich structures with the high spatial resolution of ultrasound detection~\cite{Kruger95,Kruger99,Xu11}. In PACT, the biological tissue of interest is irradiated with a short laser pulse. Under the condition of thermal confinement, the absorption of the optical energy results in the generation of pressure waves via the photoacoustic effect~\cite{Oraevsky03,Gusev93}. These pressure waves are subsequently detected by use of broadband ultrasound transducers. The image reconstruction problem in PACT is to estimate the absorbed optical energy density from the measured photoacoustic signals. Such an image may be of great importance for preclinical and clinical use, as there exists a strong correlation between optical absorption and the pathological condition of tissue~\cite{Jones80,Cheong03,Xu11}.

		The majority of PACT reconstruction methods~\cite{Xu99,Xu03} are based on idealized imaging models that assume an acoustically homogeneous medium. However, these assumptions are violated in many applications of PACT. For example, in small animal imaging applications, bone and/or gas pockets represent strong sources of acoustic heterogeneity. When the spatially variant acoustic properties of the medium are not accounted for in the imaging model, the reconstructed images  may contain significant distortions and artifacts~\cite{Jin08,XuHetero03}. However, in practice, it is challenging to estimate these acoustic properties accurately~\cite{Dean-Ben11}.

		To circumvent the need for detailed information regarding an object's acoustic properties  as well as suppress artifacts due to presence of acoustic heterogeneity, we have previously proposed a half-time-based reconstruction method~\cite{Anastasio05a}. The half-time-based reconstruction method exploits data redundancies~\cite{Anastasio05} to uniquely and stably reconstruct images from measurement data that are uniformly truncated with respect to the temporal coordinate. Although the  half-time-based reconstruction method mitigates acoustic heterogeneity-induced artifacts,  the reconstructed images can still contain significant distortions. Moreover, half-time-based methods do not employ any \textit{a priori} information about the location of strong acoustic heterogeneities in the reconstruction.

		In addition to truncation based strategies, work has also been conducted on incorporating statistical information about the object to mitigate artifacts due to acoustic heterogeneities~\cite{Dean-Ben11}. In that approach, \textit{a priori} information about the acoustic properties of the object are utilized to probabilistically weigh the tomographic contribution of each detector to a pixel in the reconstructed image~\cite{Dean-Ben11}.  Such a statistical approach was shown to have merit for mitigating artifacts due to weak acoustic heterogeneities. However, in the presence of strong heterogeneities, the approach has not been demonstrated to be effective~\cite{Dean-Ben11}.

		In this work, a PACT reconstruction method is proposed that is based on a variable data truncation (VDT) approach. This method represents a generalization of the half-time reconstruction method. The VDT-based reconstruction method employs \textit{a priori} information about the location of the isolated acoustic heterogeneities but does not require information regarding its acoustic properties. In the VDT-based method, the degree of temporal truncation is dependent on the location of the isolated acoustically heterogeneous region relative to the ultrasound transducer positions. Due to the adaptive nature of the temporal truncation, artifacts arising from the acoustic heterogeneities can in some cases be more effectively suppressed in images reconstructed by this method as compared to images reconstructed by the half-time-based method.

		\section{Background: imaging models and reconstruction methods for PACT}
		\label{sec:Background}
		Conventional imaging models and reconstruction methods for PACT are reviewed below. The previously proposed half-time reconstruction method~\cite{Anastasio05a,Anastasio05} for PACT is also reviewed.
		\subsection{Photoacoustic wavefield propagation: Continuous and discrete formulation}
		\label{sec:Models}
		Let $p(\mathbf{r},t)$ denote the photoacoustically-induced pressure wavefield at location $\mathbf{r} \in \mathbb{R}^3$ and time $t\geq 0$. Additionally, let $p_0(\mathbf{r})$ denote the initial pressure distribution generated within the object due to the photoacoustic effect, and $\mathbf{u}(\mathbf{r},t)$ denote the vector-valued acoustic particle velocity. In our formulation of the PACT imaging model, the object and the surrounding medium are assumed to possess homogeneous and lossless acoustic properties. Let $c_0$ denote the medium's uniform speed of sound (SOS) value, and $\rho_0$ denote the distribution of the uniform ambient density value.
		
		In a lossless acoustically homogeneous fluid medium, the propagation of $p(\mathbf{r},t)$ can be modeled by the following coupled equations~\cite{Treeby10a,Morse87}:
		\begin{subequations}\label{eq:diffeq}
			\begin{align}
			\frac{\partial}{\partial t}\mathbf{u}(\mathbf{r},t) &= - \frac{1}{\rho_0} \nabla p(\mathbf{r},t) \\
			\frac{\partial}{\partial t}p(\mathbf{r},t) &= - \rho_0 c_0^2 \nabla \cdot \mathbf{u}(\mathbf{r},t),
			\end{align}
			
			subject to the initial conditions 
			\begin{align}
			p(\mathbf{r},t)|_{t=0} = p_0(\mathbf{r}),\ \ \ \ \ \mathbf{u}(\mathbf{r},t)|_{t=0} = 0.
			\end{align}
		\end{subequations}

		Consider that $p(\mathbf{r},t)$ is recorded outside of the support of the object for $\mathbf{r} \in \text{d}\Omega$ and $t \in [0,T]$, where $\text{d}\Omega \subset \mathbb{R}^3$ denote a continuous measurement aperture. Throughout this study we will assume $\text{d}\Omega$ is a 2-D circular aperture with radius $R_0$. The circular aperture $\text{d}\Omega$ is assumed to be parallel to the x-y plane and is located a distance $z'$ away from the plane $z=0$. In this case, the imaging model can be described as a continuous-to-continuous (C-C) mapping as:
		\begin{align}\label{eq:C-C}
		p(\mathbf{r},t) = \mathcal{M} \mathcal{H}p_0(\mathbf{r}),
		\end{align}
		where $\mathcal{H} : \mathbb{L}^2(\Omega) \mapsto \mathbb{L}^2(\Omega \times [0,T])$ is a linear operator that denotes the action of the wave equation given by Eqn.~\eqref{eq:diffeq}, $p(\mathbf{r},t) \in \mathbb{L}^2(\text{d}\Omega\times [0,T])$ denotes the measured data function and the operator $\mathcal{M}$ is the restriction of $\mathcal{H}$ to $\text{d}\Omega \times [0,T]$.

		In practice, the detected pressure wavefield is discretized temporally and spatially at specific transducer locations. Let $\mathbf{p} \in \mathbb{R}^{N_rL}$ denote the discretized pressure signal. Unless stated otherwise, throughout the study we shall neglect the effects of the acousto-electric impulse response (EIR) as well as the spatial impulse response (SIR) of the transducer~\cite{Kenji14,Wang12,Turner14}. A continous-to-discrete (C-D) PACT imaging model~\cite{Wang11,Barrett04} can be generally expressed as 
		\begin{align}\label{eq:C-D}
		[\mathbf{p}]_{kL + l}= p(\mathbf{r},t )|_{\mathbf{r} = \mathbf{r}_0^k,t=l\Delta^t}
		\end{align}
		for  $k = 0,1,2,...,N_r-1 \text{ and } l = 0,1,2,...,L-1$. Here, $L$ is the total number of temporal samples, $\Delta^t$ is the temporal sampling interval, and the vectors $\mathbf{r}_0^k \in \mathbb{R}^3,\ k = 0,1,2,....,N_r-1$, represent the position vectors of the $N_r$ receivers on the aperture $\text{d}\Omega$.
		
		To obtain a discrete-to-discrete (D-D) imaging model for use in numerically simulating PA wavefield propagation, a finite-dimensional approximate representation of the object function $p_0(\mathbf{r})$ needs to be introduced. Also, we will assume that the reconstruction method estimates a 2-D slice of the object in plane with the circular aperture $\text{d}\Omega$. Thus, the finite-dimensional representation of 2-D slice of the object function $p(\mathbf{r})$ is given by
		\begin{align}
		p_0^{a} (\mathbf{r}',z') = \sum_{n=0}^{N-1} [\boldsymbol{\theta}]_{n}\phi_{n}(\mathbf{r}'),
		\end{align}
		where $\mathbf{r}' = (x,y)^T$ and $\{\phi_n(\mathbf{r}')\}_{{n=0}}^{{N-1}}$ are pixel expansion functions defined as
		\begin{align}\label{eq:expansion}
		\phi_n(\mathbf{r}') = 
		\begin{cases}
		1,\ \text{  if  } |x-x_n| \leq \frac{\Delta x}{2} \text{ and } |y - y_n| \leq \frac{\Delta y}{2}, \\
		0,\ \text{  otherwise}
		\end{cases}.
		\end{align} 
		Here, $(x_n,y_n)^T$  specifies the coordinate of the $n^{th}$ grid point of a uniform 2-D Cartesian with N grid points in plane with $\text{d}\Omega $. Furthermore, $\Delta x$ and $\Delta y$ represents the uniform grid spacing in the x- and y-direction, respectively. Note, that the above described finite-dimensional representation of the object function $p(\mathbf{r})$ is based on the pixel based discretization approach. However, the representation of the object function can be generalized to other basis functions such as spherical voxels~\cite{Wang14}, etc.

		Thus, given $\boldsymbol{\theta}$, $c_0$, and $\rho_0$, a D-D imaging model is given by
		\begin{align}
		\label{eq:DDmodel}
		\mathbf{p} = \mathbb{M} \mathbb{H}\boldsymbol{\theta},
		\end{align}  
		where $\mathbb{H} \in \mathbb{R}^{NL \times N}$ is the discrete approximation of the wave operator $\mathcal{H}$ that solves the initial value problem defined in Eqn.~\eqref{eq:diffeq}. Here, $\mathbb{M} \in \mathbb{R}^{N_rL \times NL}$ is a sampling matrix that models the PACT data acquisition process. For simplicity, we assume that the transducers are point-like in this study. When the receiver and grid point locations do not coincide, an interpolation method is required. Hence, the elements of $\mathbb{M}$ are chosen such that Delaunay triangulation~\cite{Lee80} based interpolation is performed.  The goal of image
		reconstruction in a discrete setting is to determine an estimate of $\boldsymbol{\theta}$ by use
		the measured data $\mathbf{p}$.
		
		\subsection{Full-time based reconstruction methods}
		In this section, the salient features of the full-time-based backprojection (BP) method and the full-time-based iterative reconstruction algorithm are discussed. In the full-time-based methods, the images are reconstructed from full-time data. The full-time data refers to the data recorded at all transducers for time $t_{full}$, where $t_{full} = 2\times \text{max}(t^1,t^2,...,t^{N_r})$. Here, $t^k$ is the time it takes for pressure waves to propagate from the center of the circular measurement aperture $\text{d}\Omega$ to the $k^{th}$ transducer in an acoustically homogeneous medium, where $\mathbf{r}_0^k \in \text{d}\Omega$.
		\subsubsection{Backprojection method}
		A variety of backprojection-type PACT image reconstruction methods\cite{Kunyansky07,Finch04,Finch07,Burghozler07} have been developed based on the continuous imaging model described by Eqn.~\eqref{eq:C-C}. In the presence of discretization, these methods provide an approximate estimate of the object function. The finite-dimensional estimate of the object function obtained by use of the BP method is given by~\cite{Xu05} 
		\begin{equation}\label{eq:BP}
		[\hat{\boldsymbol{\theta}}]_m = \sum_{k=0}^{N_r-1} {\Delta\Omega_k B\left(\mathbf{p},\frac{|\mathbf{r}_0^k- \mathbf{r}_m|}{c_0 \Delta^t},k\right)}\Big/ \sum_{k=0}^{N_r-1} {\Delta \Omega_k},
		\end{equation}
		where $[\hat{\boldsymbol{\theta}}]_m$ is the $m^{th}$ element of $\hat{\boldsymbol{\theta}}$ corresponding to the grid position $\mathbf{r}_m = (x_m,y_m,z')^T$ and $\Delta\Omega_k$ is computed as
		\begin{equation}\label{eq:sangle}
		\Delta\Omega_k = \frac{\Delta S_k}{|\mathbf{r}_m- \mathbf{r}_0^k|^2}\left[ \mathbf{n}_{0k}^S \cdot \frac{\mathbf{r}_m - \mathbf{r}_0^k}{|\mathbf{r}_m- \mathbf{r}_0^k|} \right].
		\end{equation}
		In Eqn. \eqref{eq:sangle}, $\Delta S_k$ represents the surface area occupied by the transducer at position $\mathbf{r}_0^k$, and $\mathbf{n}_{0k}^S$ is the normal of the measurement surface at $\mathbf{r}_0^k$ that points towards the photoacoustic source.
		The function $B$ in Eqn. \eqref{eq:BP} is a linear interpolation function that is defined as 
		\begin{align}
		B(\mathbf{p},t_{val},k) = &(t_{val}- t_{int})[\mathbf{p}]_{kL + (t_{int}+1)} \nonumber \\
		&+ (t_{int}+1-t_{val}) [\mathbf{p}]_{kL + t_{int}},
		\end{align}
		where $t_{int} = \lfloor t_{val}\rfloor $. Compared to the universal BP formula derived by Xu and Wang~\cite{Xu05}, the backprojection term $B\left(\mathbf{p},\frac{|\mathbf{r}_0^k-  \mathbf{r}_m|}{c_0 \Delta^t},k\right)$ in Eqn.~\eqref{eq:BP} does not contain the temporal derivative of pressure. This allows us to mitigate the impact of noise in the reconstructed image as the derivative term amplifies the contribution of noise in the measured data to the reconstructed image. Moreover, the BP formula in Eqn.~\eqref{eq:BP} assumes a full spherical detector surface~\cite{Xu05}. Since the detector geometry used in our applications are circular, the BP reconstruction formula presented in Eqn.~\eqref{eq:BP} is only approximately applicable. Nevertheless, the reconstruction formula presented in Eqn.~\eqref{eq:BP} has been used empirically to yield reconstructed images of the initial pressure distribution~\cite{Nasiriavanaki13,Yao14}.  However, quantitatively accurate 2D filtered BP reconstruction formula for circular detector geometry have  been established~\cite{Burghozler07,Finch07}. 
		\subsubsection{Optimization-based image reconstruction algorithms}
		Since BP-type reconstruction methods are based on the discretization of an analytical reconstruction formula, they possess several limitations. For instance, BP reconstruction methods require the measured data to be densely sampled on an aperture that encloses the object.
		% This proves problematic as acquiring densely sampled acoustic measurements can require expensive transducer arrays and/or long data-acquisition times if mechanical scanning is employed.
		Moreover, analytical reconstruction methods ignore measurement noise.
		Hence, they can yield reconstructed images that have sub-optimal trade-offs between image variance and spatial resolution. The use of iterative PACT image reconstruction algorithms can circumvent these shortcomings~\cite{Wang12}.
		% Iterative image reconstruction algorithms can improve image quality in the presence of measurement noise and permit reductions in data-acquisition times as compared with those yielded by BP reconstruction methods~\cite{Wang12}.  

		Commonly employed iterative PACT reconstruction algorithms 
		seek to compute penalized least squares (PLS) estimates  by solving an optimization problem of the form
		\begin{align}\label{eq:optim}
		\hat{\boldsymbol{\theta}} = \underset{\boldsymbol{\theta}}{\operatorname{argmin}} {||\mathbf{p} - \mathbb{M} \mathbb{H} \boldsymbol{\theta}||^2_2} + \lambda R(\boldsymbol{\theta}),
		\end{align}
		where $\lambda$ is a regularization parameter and $R(\theta)$ is a regularizing penalty term. For this study, the total variance (TV) semi-norm penalty, given by
		\begin{align}
		R(\boldsymbol{\theta}) = ||\boldsymbol{\theta}||_{TV} \equiv \sum_{n=1}^N \Big\{ ([\boldsymbol{\theta}]_n - [\boldsymbol{\theta}]_{n1^{-}})^2 + ([\boldsymbol{\theta}]_n - [\boldsymbol{\theta}]_{n2^{-}})^2\Big\}^{\frac{1}{2}},
		\end{align}
		was employed. Here, $[\boldsymbol{\theta}]_n$ denotes the $n^{th}$ grid node, and $[\boldsymbol{\theta}]_{n1^{-}}, [\boldsymbol{\theta}]_{n2^{-}}$ are the neighboring nodes before the $n^{th}$ node along the first and second dimension, respectively.
		\subsection{Half-time based reconstruction methods}
		\label{sec:Half}
		It has been shown that images reconstructed using full-time data can contain significant distortions when acoustic variations in the object are not accurately modeled~\cite{Anastasio05,XuHetero03,Modgil10}. To address this problem, an image reconstruction method based on a data truncation strategy known as the half-time method~\cite{Anastasio05a,Anastasio05} was previously proposed. In the half-time method, all transducer measurements are temporally truncated at a specific delay time $t_{half}$, where $t_{half} = \frac{t_{full}}{2}$. For a circular measurement aperture, it can be shown that $t_{half} = \frac{R_0}{c_0}$.

		In half-time based reconstruction methods, the data vector $\mathbf{p}$ is truncated at a constant delay time $t_{half}$. This truncation process can be described by a truncation matrix $\mathbb{T}_{HALF} \in \mathbb{R}^{N_rL \times N_rL}$ that acts on $\mathbf{p}$ to produce a truncated data vector $\mathbf{p}_{\text{Trunk}} \in \mathbb{R}^{N_rL}$ as  
		\begin{align}\label{eq:trunkbp}
		\mathbf{p}_{\text{Trunk}} = \mathbb{T}_{HALF} \mathbf{p}.
		\end{align}
		Details regarding the construction of truncation matrix for the half-time method are described next.

		Define the matrix $\mathbf{T}^{m}_{half} \in \mathbb{R}^{L \times L}$ as
		\begin{align}
		[\mathbf{T}^{m}_{half}]_{ij} = 
		\begin{cases}
		1,& \text{if  } i=j \text{ and } i\Delta^t \leq \ \frac{R_0}{c_0}\\
		0,& \text{otherwise}
		\end{cases},
		\end{align}
		where $i=0,1,...,L-1$, $j=0,1,...,L-1$, and $m= 0,1,...,N_r-1$. The truncation matrix $\mathbb{T}_{HALF}$ is formed as 
		\begin{align}\label{eq:THALF}
		\mathbb{T}_{HALF} \ \ =  
		\begin{bmatrix}
		&\mathbf{T}^{0}_{half} \ &\mathbf{0_{L \times L}} &\cdots &\cdots &\mathbf{0_{L \times L}}\\
		&\mathbf{0_{L \times L}} &\mathbf{T}^{1}_{half} &\mathbf{0_{L \times L}} &\cdots  &\mathbf{0_{L \times L}}\\
		&\vdots &\mathbf{0_{L \times L}} &\ddots &\ddots &\mathbf{0_{L \times L}}\\
		&\vdots &\vdots &\ddots &\ddots &\vdots\\
		&\mathbf{0_{L \times L}}\ &\mathbf{0_{L \times L}} &\cdots &\cdots &\mathbf{T}^{N_r-1}_{half}\\
		\end{bmatrix},
		\end{align}
		where $\mathbf{0_{L \times L}}$ is the L $\times$ L zero matrix.

		By replacing the full-time data vector with a half-time data vector in a full-time reconstruction method, a half-time reconstruction method is readily obtained. For example, a half-time BP method is expressed as  
		\begin{equation}\label{eq:BPhalftrunk}
		[\hat{\boldsymbol{\theta}}]_m = \sum_{k=0}^{N_r-1} {\Delta\Omega_k B\left(\mathbb{T}_{HALF}\mathbf{p},\frac{|\mathbf{r}_0^k- \mathbf{r}_m|}{c_0 \Delta^t},k\right)}\Big/ \sum_{k=0}^{N_r-1} {\Delta \Omega_k}.
		\end{equation}
		Similarly, a half-time based iterative algorithm can be expressed as 
		\begin{equation}\label{eq:optimhalftrunk}
		\hat{\boldsymbol{\theta}} = \underset{\boldsymbol{\theta}}{\operatorname{argmin}} {||\mathbb{T}_{HALF}\mathbf{p} - \mathbb{T}_{HALF}\mathbb{M} \mathbb{H} \boldsymbol{\theta}||_2}^2 + \lambda ||\boldsymbol{\theta}||_{TV}.
		\end{equation}
		
		\section{Image reconstruction based on variable truncation methods}
		Even though half-time-based reconstruction methods mitigate acoustic heterogeneity-induced artifacts, the reconstructed image can, in some cases still contain significant artifacts. In scenarios where an acoustically heterogeneous region is localized and its support is relatively small compared to the area of the reconstruction region, the half-time truncation method can be readily extended to a variable data truncation (VDT) method. Unlike the half-time method, the temporal truncation strategy employed in the VDT method utilizes \textit{a priori} information about the location of the acoustically heterogeneous region. The VDT method is applicable when the acoustic heterogeneity is isolated in location from the features that are likely to be imaged and when the support of the acoustically heterogeneous region is small relative to the area of the reconstruction region. In small animal imaging applications, bone, spinal column, and gas pockets represent isolated acoustic heterogeneities.  

		\subsection{Geometrical interpretation}
		In the VDT method, the measurement data recorded at each transducer are temporally truncated based on the distance between the corresponding transducer and the nearest isolated acoustic heterogeneity. As a result, all data that have been strongly influenced by an acoustic heterogeneity are discarded and not employed for image reconstruction.

\begin{figure}[!t]
			\centering
			{\includegraphics[height=5.5cm]{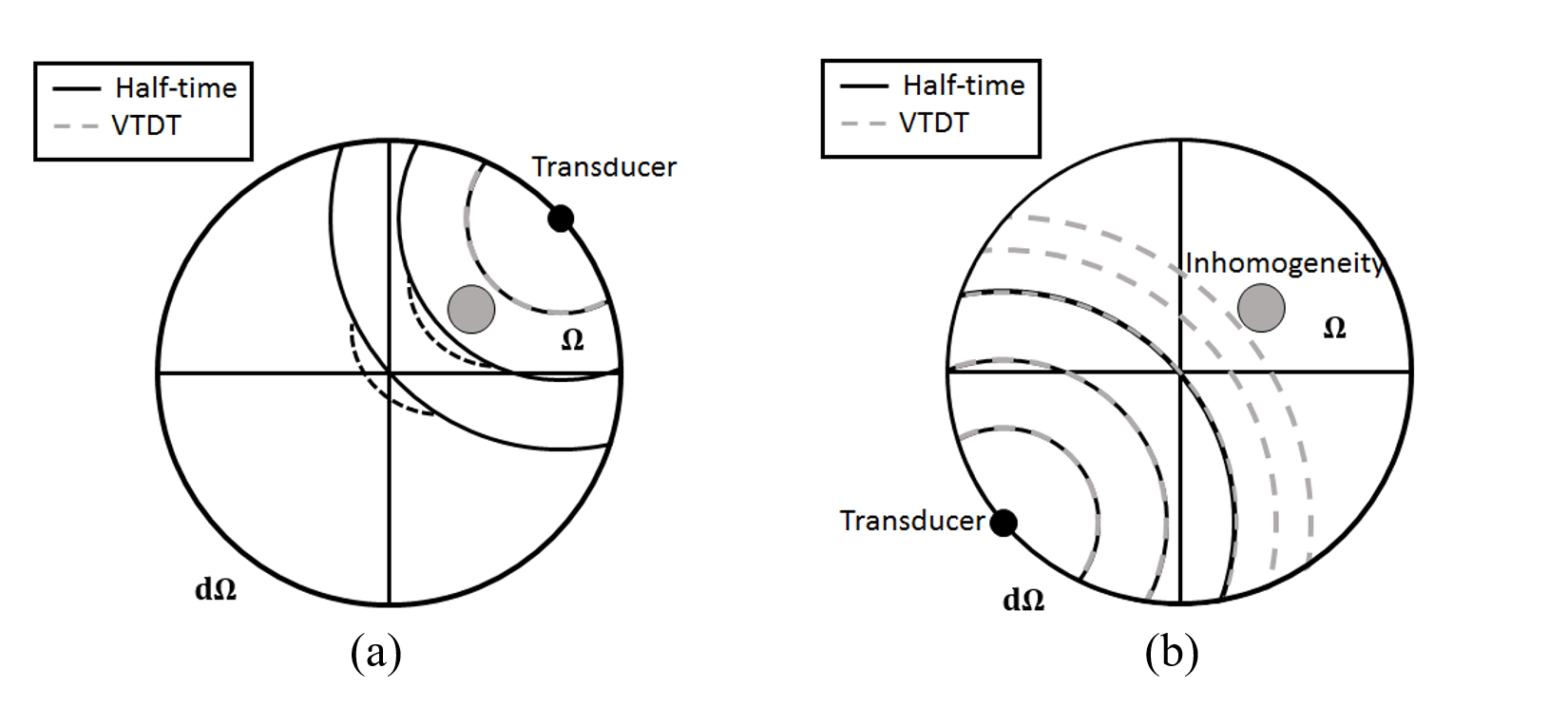}%
			}
			\caption{A schematic showing the iso-time of flight contours for the half-time-based and the VDT-based method for two transducers located opposite one another. (a) The half-time-based BP method backprojects data beyond the acoustic heterogeneity. The distorted wavefront is denoted by the black dashed line. For the purpose of this diagram, we assume that the background speed of sound of the medium is less than the speed in the acoustically heterogeneous region. (b) The half-time method truncates the data prior to encountering any acoustic heterogeneity. }
			\label{Fig:isoTOF}
		\end{figure}

		As depicted in Fig.~\ref{Fig:isoTOF}a, the pressure signals that have been distorted by traveling through the acoustic heterogeneity are not truncated in the half-time measurement data. When a reconstruction method employing such data does not account for this distortion, it can be a source of significant artifacts in the reconstructed image. The VDT measurement data, on the other hand, do not contain pressure signals that are reflected by or transmitted through the acoustic heterogeneity (see Fig.~\ref{Fig:isoTOF}b). Hence, the artifacts introduced by these distorted pressure signals are eliminated in the images reconstructed using VDT-based methods. Additionally, for the transducer shown in Fig.~\ref{Fig:isoTOF}b, the VDT-based method uses more temporal data than the half-time-based reconstruction method. In the subsequent section, we will describe the construction of the truncation matrix for the VDT method. 
		\subsection{Construction of VDT truncation matrix}
		\label{sec:VDTM}
\begin{figure}[!t]
			\centering
			{\includegraphics[height = 2.0in, width=2.5in]{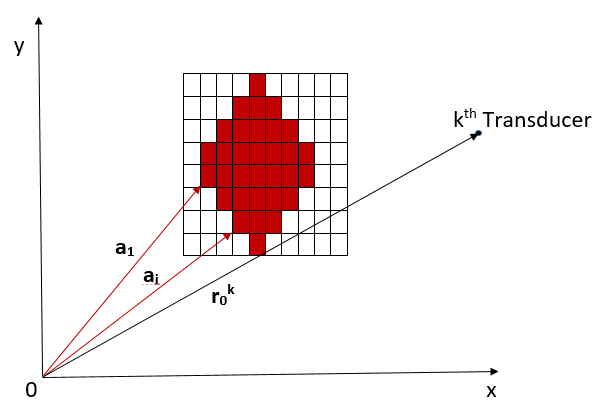}%
				\caption{The red pixels represent the location of acoustic heterogeneities in the 2-D cartesian grid. The collection of position vectors $\mathbf{a}_1,..,\mathbf{a}_i,..,\mathbf{a}_h$ form the set $\mathbb{A}_h$. The position vector of the $k^{th}$ transducer, $\mathbf{r}_0^k$ is also shown.
					\label{Fig:Grid}}}
		\end{figure}

		Similar to the half-time based method, the truncation process in the VDT-based method can be described by a truncation matrix $\mathbb{T}_{VDT} \in \mathbb{R}^{N_rL \times N_rL}$ that acts on the data vector $\mathbf{p}$. Thus, replacing the truncation matrix in Eqn.~\eqref{eq:trunkbp} with $\mathbb{T}_{VDT}$, we have 
		\begin{align}\label{eq:trunckVDT}
		\mathbf{p}_{\text{Trunk}} = \mathbb{T}_{VDT} \mathbf{p}.
		\end{align}
		Let  $\mathbf{a}_h \in \mathbb{R}^2, \text{ for  }h = 0,1,..., N_h-1 $, denote the position of a grid point where the acoustic properties differ from the background values. Let $\mathbb{A}_h$ denote the collection of all such points (see Fig.~\ref{Fig:Grid}). The nearest grid point to the $k^{th}$ transducer at $\mathbf{r}_0^k$ that corresponds to an acoustic heterogeneity is given by 
		\begin{equation}
		\mathbf{a_T}^k = \underset{\mathbf{a}_h \in \mathbb{A}_h}{\operatorname{argmin}} ||\mathbf{a}_h - \mathbf{r}_0^k||_2 .
		\end{equation}
		
		Define the matrix $\mathbf{T}^{m}_{VDT} \in \mathbb{R}^{L \times L}$ as  
		\begin{align}
		[\mathbf{T}^{m}_{VDT}]_{ij} = 
		\begin{cases}
		1,& \text{if  } i=j \text{ and } i\Delta^t \leq \ \frac{||\mathbf{a_T}^m - \mathbf{r}_0^m||_2}{c_0} \\
		0,& \text{otherwise}
		\end{cases},
		\end{align}
		where $i=0,1,...,L-1$, $j=0,1,...,L-1$ and $m= 0,1,...,N_r-1$. Hence, the truncation matrix $\mathbb{T}_{VDT}$ is given by  
		\begin{align}\label{eq:TVDT}
		\mathbb{T}_{VDT} \ \ =  
		\begin{bmatrix}
		&\mathbf{T}^{0}_{VDT} \ &\mathbf{0_{L \times L}} &\cdots &\cdots &\mathbf{0_{L \times L}}\\
		&\mathbf{0_{L \times L}} &\mathbf{T}^{1}_{VDT} &\mathbf{0_{L \times L}} &\cdots  &\mathbf{0_{L \times L}}\\
		&\vdots &\mathbf{0_{L \times L}} &\ddots &\ddots &\mathbf{0_{L \times L}}\\
		&\vdots &\vdots &\ddots &\ddots &\vdots\\
		&\mathbf{0_{L \times L}}\ &\mathbf{0_{L \times L}} &\cdots &\cdots &\mathbf{T}^{N_r-1}_{VDT}\\
		\end{bmatrix}.
		\end{align}
		As defined above, the VDT truncation matrix is dependent upon the location and size of acoustic heterogeneity. The half-time truncation matrix, on the other hand, is independent of these. As a result, the VDT truncation matrix discards all the pressure signals reflected off or transmitted through the heterogeneity, while the half-time truncation matrix may not necessarily do so.
		
		\subsection{VDT reconstruction methods}
		VDT-based reconstruction methods can be formed readily by replacing the full-time data vector in a full-time reconstruction method by its truncated version. Thus, the VDT-based BP method to reconstruct $\hat{\boldsymbol{\theta}}$ is given by 
		\begin{equation}\label{eq:BPtrunk}
		[\hat{\boldsymbol{\theta}}]_m = \sum_{k=0}^{N_r-1} {\Delta\Omega_k B\left(\mathbb{T}_{VDT}\mathbf{p},\frac{|\mathbf{r}_0^k- \mathbf{r}_m|}{c_0 \Delta^t},k\right)}\Big/ \sum_{k=0}^{N_r-1} {\Delta \Omega_k}.
		\end{equation}
		Similarly, the VDT-based iterative algorithm is given by
		\begin{equation}\label{eq:optimtrunk}
		\hat{\boldsymbol{\theta}} = \underset{\boldsymbol{\theta}}{\operatorname{argmin}} {||\mathbb{T}_{VDT}\mathbf{p} - \mathbb{T}_{VDT}\mathbb{M} \mathbb{H} \boldsymbol{\theta}||_2}^2 + \lambda ||\boldsymbol{\theta}||_{TV}.
		\end{equation}
		\section{Computer-Simulation Studies}
		\label{sec:Sim}
		
		Computer-simulation studies were conducted to compare the performance of  the VDT- and half-time-based reconstruction methods.  
		\subsection{Methods}
\begin{figure}[!t]
			\centering
				{\includegraphics[height = 2.5in,width = 6.5in]{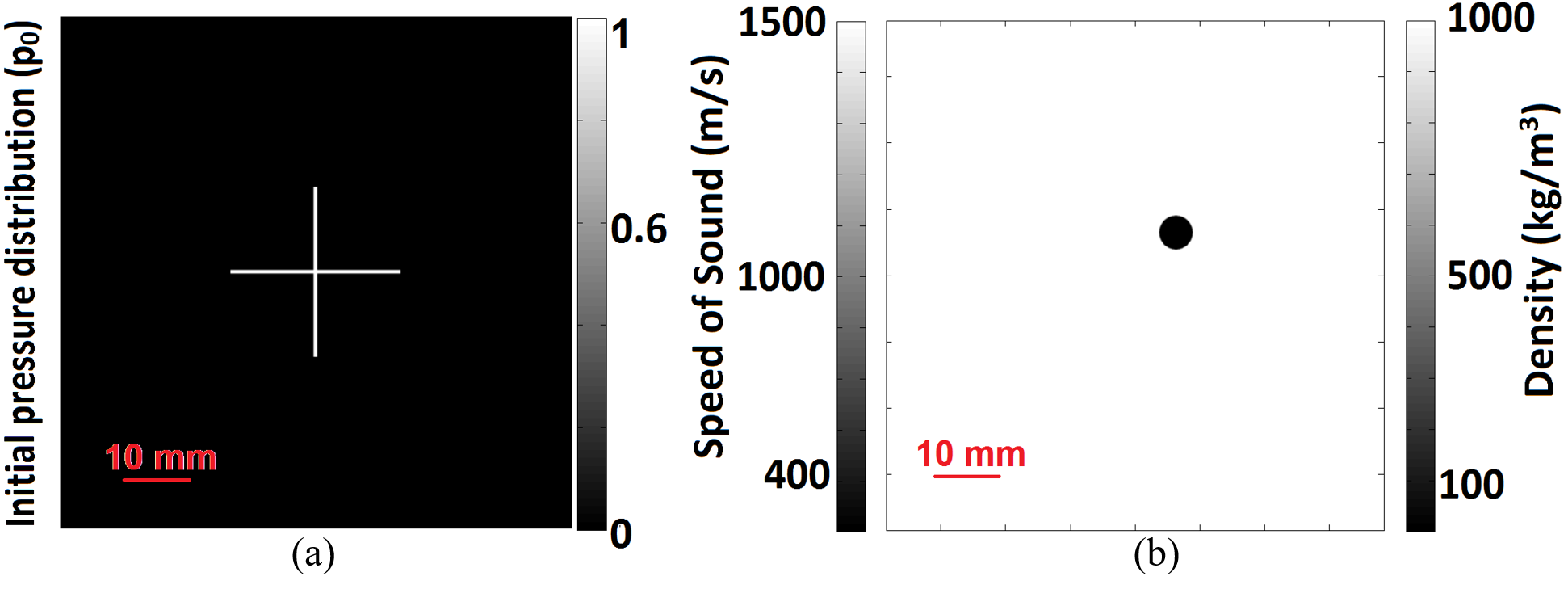}}
			\caption{(a) The initial pressure distribution, (b) speed of sound and density distributions used for computer simulation studies.}
			\label{Fig:Figure1}
			
		\end{figure}

		The k-space pseudospectral method~\cite{Mast01,Tabei02} for numerically solving the photoacoustic wave equation has been implemented in the MATLAB k-wave toolbox~\cite{Treeby10}. This toolbox was employed to compute the action of forward operator $\mathbb{H}$.  
		A 2-D circular scanning geometry consisting of 512 transducers that were evenly distributed in a circle of radius 50 mm was employed. 
		The numerical phantoms employed in this study to generate forward data are shown in Fig.~\ref{Fig:Figure1}. The acoustic heterogeneity employed in the simulation, shown in Fig.~\ref{Fig:Figure1}b, imitated an air void ($c_0$ = 340 $\frac{m}{s}$ and $\rho_0$ = 1.2 $\frac{kg}{m^3}$) while the background medium consisted of water ($c_0$ = 1500 $\frac{m}{s}$ and $\rho_0$ = 1000 $\frac{kg}{m3}$). The initial pressure phantom, in Fig.~\ref{Fig:Figure1}a consisted of two line absorbers placed perpendicular to one another. 
		operator $\mathbb{H}$. 

		Assuming ideal point-like transducers, the simulated pressure data corresponding to the numerical phantoms were calculated using the k-space pseudospectral method for the scanning geometry described. A 1536 $\times$ 1536 grid with a pitch of 500 $\mu m$ was employed to simulate the pressure data. A total of 5200 temporal samples were computed at each transducer location with a time step of $\Delta^t = 6.25\ ns$. In addition, the generated forward data on the sensors was contaminated with additive white gaussian noise to achieve a signal-to-noise ratio of 10. 
		
		For both the iterative and BP reconstruction methods, a constant speed of sound of $c_0 = 1500 \frac{m}{s}$ was assumed. For the VDT-based reconstruction methods, based on the known location of the acoustic heterogeneity and the assumed $c_0$, the truncation matrix $\mathbb{T}_{VDT}$ was established as prescribed by Eqn.~\eqref{eq:TVDT}. The truncation matrix was utilized to compute the truncated data vector $ \mathbf{p}_{\text{Trunk}}$. Similarly, the truncation matrix and the truncated pressure data vector were also computed for use with the half-time-based reconstruction methods. Furthermore, the fast iterative shrinkage/thresholding method (FISTA) ~\cite{Beck09,Huang13} was implemented to solve the optimization problems in Eqns.~\eqref{eq:optimhalftrunk} and~\eqref{eq:optimtrunk}.
		\subsection{Reconstructed Images}
\begin{figure}[!t]
			\centering
			{\includegraphics[height = 5.0in, width=5.5in]{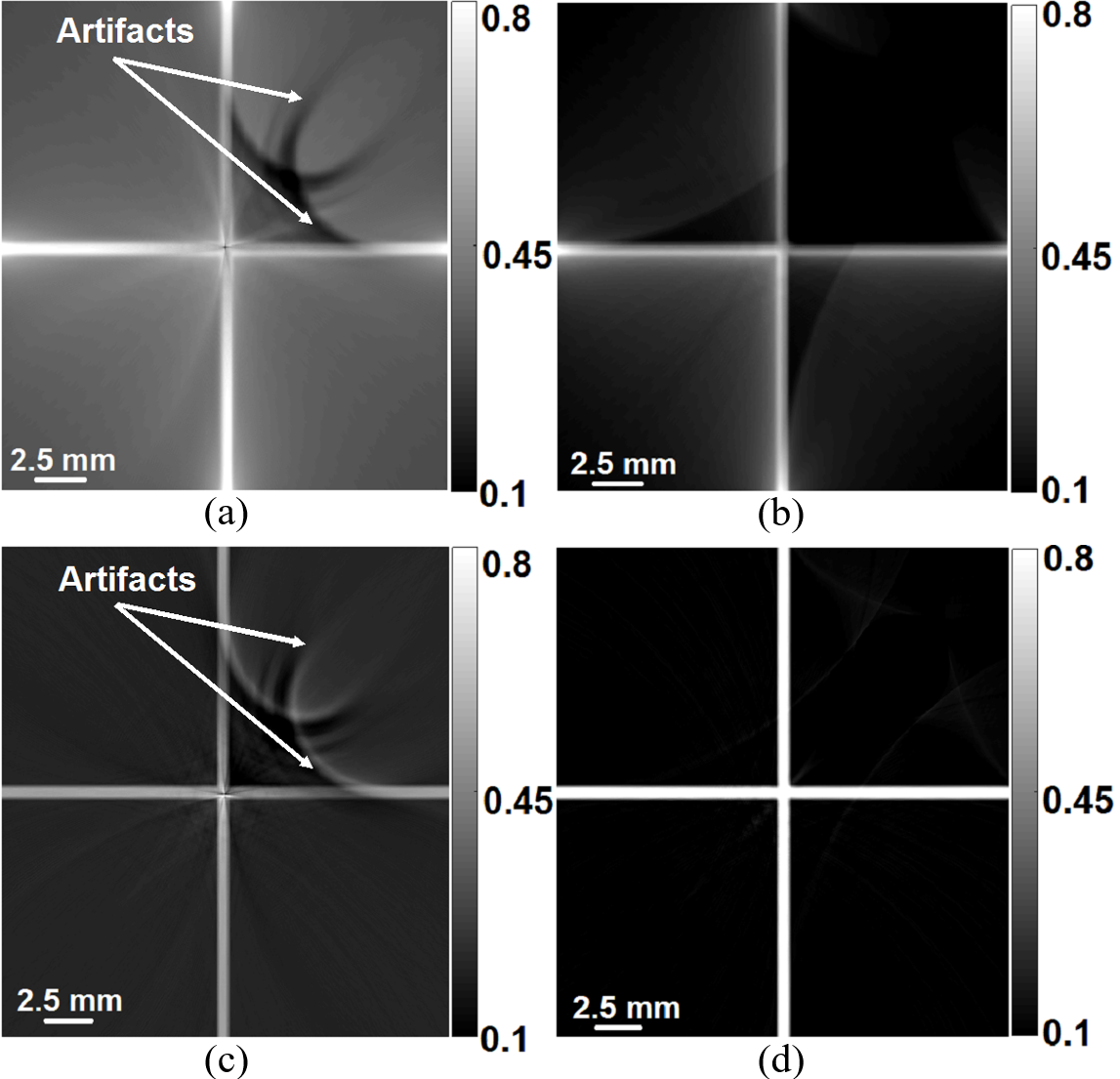}}
		\caption{Images reconstructed from the simulated forward data using (a) the half-time-based BP method, (b) the VDT-based BP method, (c) the half-time-based iterative algorithm, and (d) the VDT-based iterative algorithm.}
		\label{Fig:Figurenum}
		\end{figure}

		The images reconstructed using half-time- and VDT-based methods are shown in Fig.~\ref{Fig:Figurenum}.
		In the images reconstructed by use of the half-time-based methods, the artifacts due to the acoustic heterogeneity (air void) are marked by white arrows in Figs.~\ref{Fig:Figurenum}a and~\ref{Fig:Figurenum}c.  The same artifacts, however, are mitigated in the images reconstructed using VDT-based methods, which are shown in Figs.~\ref{Fig:Figurenum}b and~\ref{Fig:Figurenum}d. In addition, differences exists between the images reconstructed using BP and iterative methods. This difference is due to the action of the TV-penalty term in the the iterative algorithm which smooths the image while preserving its edges. The mitigation of the artifacts due to the air void can be quantified by calculating the root-mean-squared error (RMSE) between the reconstructed pressure distribution and the true pressure distribution. The RMSEs between the reconstructed pressure distribution and the original pressure distribution are shown in Table~\ref{tab:RMSE}. The RMSE values for the images reconstructed using VDT-based method are lower than  the RMSE values for the images reconstructed using half-time-based methods for both BP and iterative methods. 
		\begin{table}[!t]
			\renewcommand{\arraystretch}{1.3}
			% if using array.sty, it might be a good idea to tweak the value of
			%\extrarowheight as needed to properly center the text within the cells
			\centering
			% Some packages, such as MDW tools, offer better commands for making tables
			% than the plain LaTeX2e tabular which is used here.
			\caption{Table listing the RMSE between the reconstructed pressure distribution and the initial pressure distribution}
			\label{tab:RMSE}
			\begin{tabular}{|c||c||c|}
				\hline 
				& \textbf{Half-time-based} & \textbf{VDT-based}\\
				\hline
				\textbf{BP} & 189.33 & 104.67\\
				\hline
				\textbf{FWI} & 111.40 & 54.82\\
				\hline
			\end{tabular}
		\end{table}		
		\section{Experimental Studies}
		The performances of the half-time- and VDT-based reconstruction methods were also investigated by use of experimental data. The experimental data were acquired in the Optical Imaging Laboratory at Washington University in St. Louis.
		\subsection{System}
		The 2-D PACT system employing a ring array composed of 512 transducer elements distributed in a radius 50 mm was employed in the experimental studies. A short laser pulse (DLS9050, Continuum, 5-9 ns pulse width) with a repetition rate of 50 Hz at a wavelength of 532 nm was used to irradiate a sample located in the center of the measurement system. The generated acoustic signals were detected by transducers with a center frequency of 5MHz and a bandwidth $>$ 90\%. The data recorded by the transducers were digitized at a sampling rate of 40 MHz.
		\subsection{Experimental Phantom Studies}
\begin{figure}[t]
			\centering
			{\includegraphics[height = 2.225in, width=2.225in]{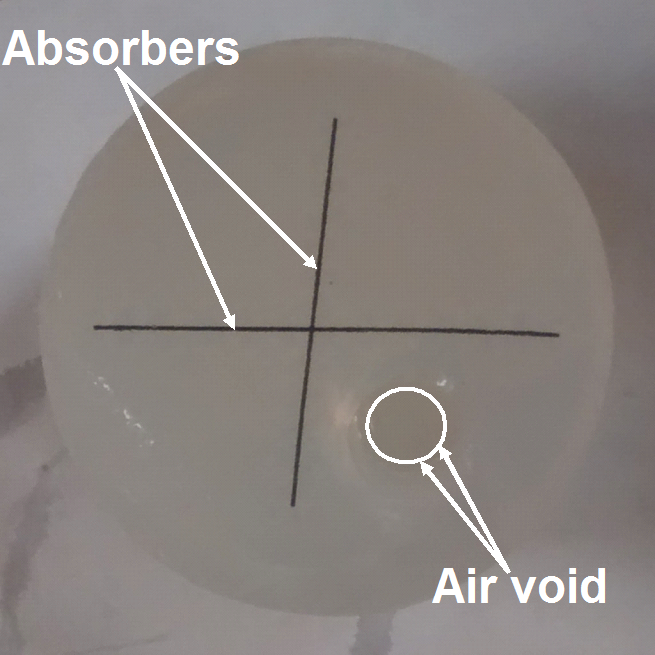}%
				\caption{An agar phantom containing two line absorbers and an air void.}
				\label{Fig:phantom}}
		\end{figure}

		The experimental agar phantom, shown in Fig.~\ref{Fig:phantom}, consisted of two linear optical absorbers (human hair) placed approximately perpendicular to one another with an air void insert in the lower right quadrant. 
		\subsubsection{Methods}
		The phantom was placed at the center of the circular transducer array and photoacoustic (PA) signals were acquired for 2000 time points with a sampling rate of 40 MHz.  Since the VDT-based reconstruction methods require \textit{a priori} information about the acoustically heterogenous region, the location of the air void relative to the transducers needed to be determined. This was accomplished through visual inspection of the image shown in Fig.~\ref{Fig:phantom}. For both the reconstruction methods, the constant speed of sound of the homogeneous background was assumed to be $1500 \frac{m}{s}$. Additionally, for the iterative reconstruction algorithm, the experimentally obtained data was pre-processed prior to reconstruction. The preprocessing of the data involved temporally up-sampling by a factor of 4 and filtering with a Hann-window low-pass filter with a cutoff frequency of 10 MHz. The preprocessing was done to avoid any issues with the numerical stability of the wave equation solver~\cite{Treeby10}. Furthermore, the fast iterative shrinkage/thresholding method (FISTA) ~\cite{Beck09,Huang13} was implemented to solve the optimization problem in Eqns.~\eqref{eq:optimhalftrunk} and~\eqref{eq:optimtrunk}.
		\subsubsection{Results}
\begin{figure}[t]
		\centering
		{\includegraphics[height = 5.0in, width=5.5in]{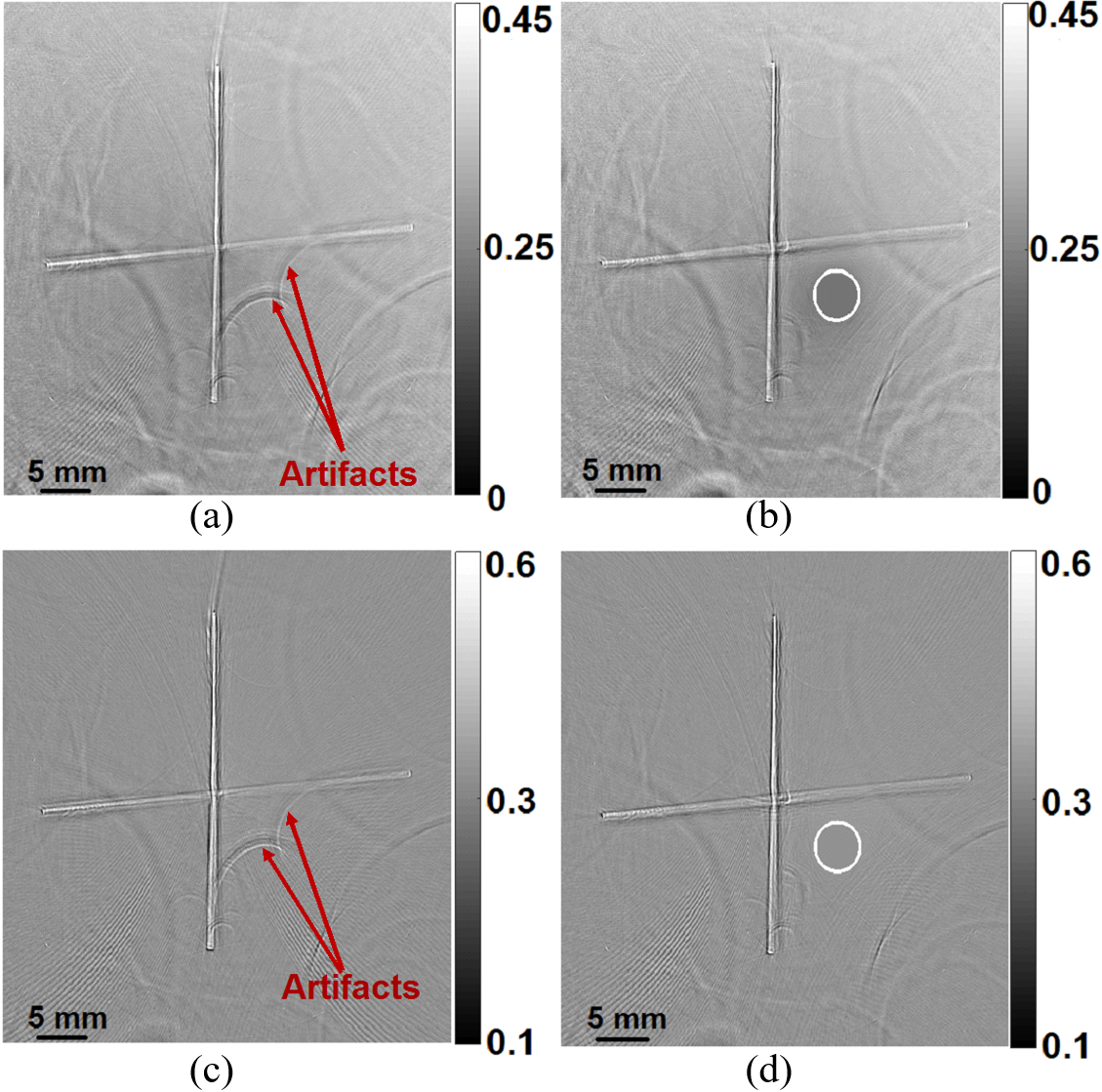}}\\
			\caption{Images reconstructed from the agar phantom measurements using (a) the half-time-based BP method, (b) the VDT-based BP method, (c) the half-time-based iterative algorithm, and (d) the VDT-based iterative algorithm. The white circle in (c) and (d) represents the boundary of the acoustic heterogeneity utilized by the VDT-based reconstruction method. All the reconstructed pressure values were mapped to the range [0,1] prior to display.}
			\label{Fig:ExptPhantom}
		\end{figure}

		Images reconstructed using the half-time- and VDT-based BP and iterative reconstruction methods are shown in Fig.~\ref{Fig:ExptPhantom}. In the images reconstructed using half-time-based approaches, the artifacts due to presence of air void are marked with red arrows. Also, the location of the acoustically heterogeneous region (air void) that was extracted from Fig.~\ref{Fig:phantom} is delineated by a white boundary as shown in Fig.~\ref{Fig:ExptPhantom}b and Fig.~\ref{Fig:ExptPhantom}d. 

		From Fig.~\ref{Fig:ExptPhantom}, the artifacts due to acoustic heterogeneity are found to be mitigated in the images reconstructed using VDT-based reconstruction methods. The artifacts are present in the lower right quadrant as shown in Figs.~\ref{Fig:ExptPhantom}a and~\ref{Fig:ExptPhantom}c.
		In addition, we also observe that the VDT-based methods are robust with regards to errors in the estimation of the boundary of the acoustically heterogeneous region. Thus, even with such inaccuracies in estimating the location and shape of air void, the VDT-based methods achieve effective mitigation of artifacts due to acoustic heterogeneities. 
		
		\subsection{Animal Studies}
		The second set of experimental data was acquired \textit{in vivo} from a mouse trunk.

		\subsubsection{Methods}
		All experimental animal procedures were carried out in conformity with laboratory animal protocols approved by the Animal Studies committee of Washington University in St. Louis. In small animal imaging applications, the major sources of acoustic heterogeneity are bones, spinal column and gas voids. These heterogeneities perturb the acoustic wavefield causing severe distortions in the reconstructed images.

		For the purposes of this study, the VDT-based reconstruction methods were employed to mitigate artifacts only due to the spinal column. In order to obtain \textit{a priori} information about the location of the acoustically heterogeneous region, we developed a semi-automatic segmentation method. The semi-automatic segmentation method estimated the boundary of the spinal column from the images reconstructed using the half-time-based BP method. As the VDT-based reconstruction methods are robust to errors in the estimation of boundary of the acoustically homogeneous region, the segmentation method did not need to be exact. The boundary of the spinal cord that was extracted by the segmentation method is depicted by the white heart-shaped contours in Figs.~\ref{Fig:Mouse}c and ~\ref{Fig:Mouse}d. We can observe that the segmentation method overestimated the area of the spinal column. Thus, all of the pressure signals reflected off or transmitted through the spinal column are truncated. 
		
		For both methods, the constant speed of sound of the homogeneous background was assumed to be $1515 \frac{m}{s}$. In order to select the optimal speed of sound value of the background, we scanned over a range of values and compared the image quality of the images reconstructed using the half-time-based BP method. The value that gave the best image quality was picked as the optimal speed of sound of the homogenous background. Similar to the experimental agar phantom data set, the data obtained from the mouse trunk were preprocessed prior to applying the iterative reconstruction algorithm. The preprocessing involved temporally up-sampling by a factor of 4 and filtering using a Hann-window low-pass filter with a cutoff frequency of 10 MHz. Furthermore, the fast iterative shrinkage/thresholding method (FISTA) ~\cite{Beck09,Huang13} was implemented to solve the optimization problems in Eqns.~\eqref{eq:optimhalftrunk} and~\eqref{eq:optimtrunk}.
		\subsubsection{Results}
\begin{figure}[t]
			\centering
			{\includegraphics[height = 5.0in,width=5.5in]{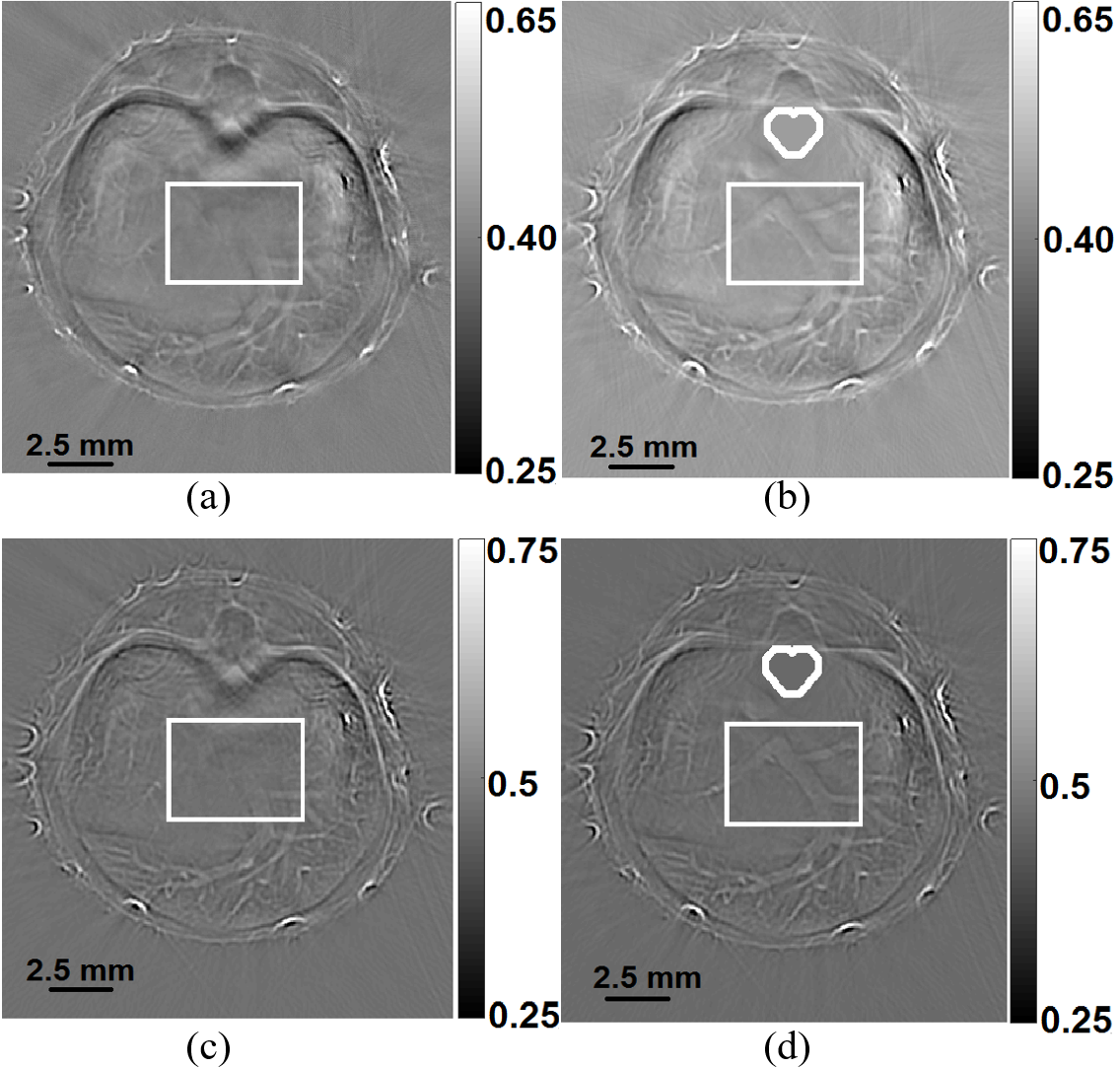}%
				\label{Fig:MouseBPHalf}}
			\caption{Images reconstructed from \textit{in vivo} mouse trunk measurements using (a) the half-time-based BP method, (b) the VDT-based BP method, (c) the half-time-based iterative algorithm, and (d) the VDT-based iterative algorithm. The heart-shaped white contour circle in (c) and (d) represents the boundary of the acoustic heterogeneity utilized by the VDT-based reconstruction methods. The region of interest delineated by white rectangular boxes in (a)-(d) are shown in great detail in Fig.~\ref{Fig:zoom}. All the reconstructed pressure values were mapped to the range [0,1] prior to display.}
			\label{Fig:Mouse}
		\end{figure}

\begin{figure}[t]
			\centering
			{\includegraphics[height = 5.0in,width=5.5in]{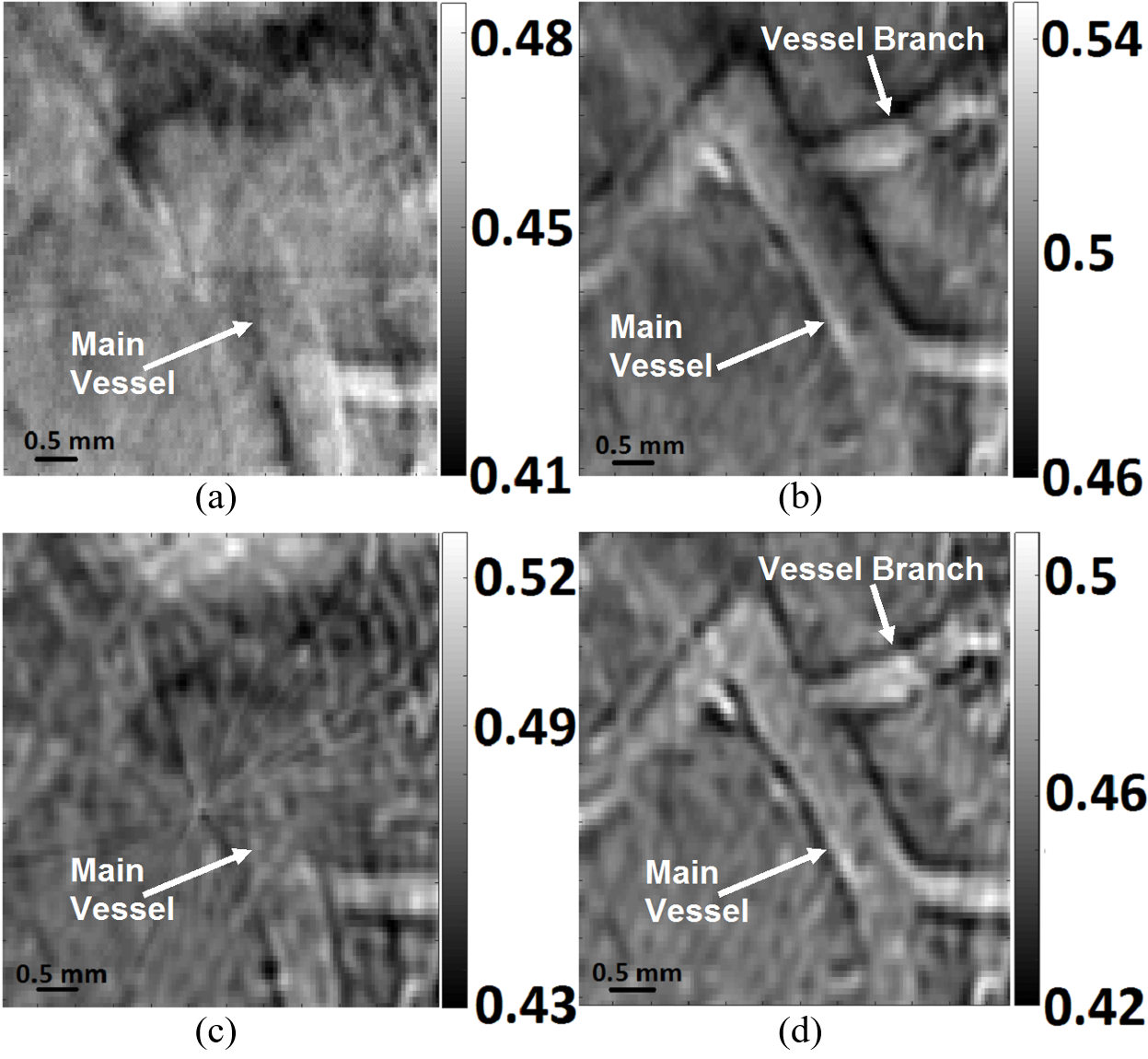}%
				\label{Fig:zoomBPhalf}}
			\caption{ The zoomed-in images for the delineated regions in the mouse-trunk images reconstucted using (a) the half-time-based BP method, (b) the VDT-based BP method, (c) the half-time-based iterative algorithm, and (d) the VDT-based iterative algorithm.}
			\label{Fig:zoom}
		\end{figure}

		Images reconstructed using the half-time- and VDT-based BP and iterative reconstruction methods are shown in Fig.~\ref{Fig:Mouse}. 
		The acoustically heterogeneous region is delineated by heart-shaped contours in the images reconstructed using VDT-based methods. To compare the performance of the VDT- and the half-time-based reconstruction methods, we analyzed the differences in the regions enclosed by white rectangular boxes in Fig.~\ref{Fig:Mouse}. The zoomed-in delineated regions of Fig.~\ref{Fig:Mouse} are shown in Fig.~\ref{Fig:zoom}. 

		In Fig.~\ref{Fig:zoom}, vessels of interest are marked by white arrows. In the images reconstructed using half-time-based methods, even though we observe the bifurcation of the lower part of the main vessel, the top part of the main vessel is obscured (see Figs.~\ref{Fig:zoom}a and ~\ref{Fig:zoom}c). However, in images reconstructed using VDT-based reconstruction methods, as shown in Figs.~\ref{Fig:zoom}b and ~\ref{Fig:zoom}d, in addition to the bifurcation of the main vessel, the top part of the vessel along with a vessel branching off from it are visible. The vessel branch as well as the main vessel are marked by red arrows in Figs.~\ref{Fig:zoom}b and ~\ref{Fig:zoom}d. 
		
		When using half-time-based reconstruction methods, the pressure signals reflected off and transmitted through the acoustic heterogeneity (spinal column) can interfere  with PA signals coming from the blood vessels, ultimately obscuring the blood vessels in the reconstructed images. However, with the VDT-based reconstruction methods, the pressure signals reflected or transmitted through the acoustic heterogeneity are truncated. Thus, certain features obscured in the images reconstructed using half-time-based methods can be clearly seen in the images reconstructed using VDT-based methods.

		\section{Conclusion}
		A VDT approach to PACT image reconstruction was proposed and investigated. The performance of VDT-based and half-time-based reconstruction methods were compared by use of simulated and experimental data sets. The PACT reconstruction methods employed, namely iterative and BP methods, were modified to include data truncation strategies in their formulation. From the results, the artifacts due to acoustic heterogeneities were mitigated much more effectively in images reconstructed using VDT-based methods as compared to the images reconstructed using half-time-based methods. The improvement in image quality was significant in the reconstructed mouse trunk images, where we found that structures obscured in the images reconstructed using half-time-based methods were visible in the images reconstructed using VDT-based methods.

		However, if the size of the heterogeneity is large or if there are multiple heterogeneities, the VDT-based approach can lead to excessive temporal truncation of the data. For such cases, the VDT-based methods may not perform better than the half-time-based methods. The problem of multiple heterogeneities represents a topic for further study. 
		
		\section*{Acknowledgment}

		This work was supported in part by NIH awards CA1744601,  EB01696301, and 5T32EB01485505.

%%%%% References %%%%%
%
\newpage
\bibliography{report}   % bibliography data in report.bib
\bibliographystyle{spiejour}   % makes bibtex use spiejour.bst
%
%%%%%% Biographies of authors %%%%%
%
%\vspace{2ex}\noindent\textbf{First Author} is an doctoral student at Washington University in St. Louis. He received his Bachelor in Applied Science degree from Simon Fraser University. His current research interests include image reconstruction algorithms in photoacoustic tomography. He is a member of SPIE.
%
%\vspace{1ex}
%\noindent Biographies and photographs of the other authors are not available.

\listoffigures
\listoftables

\end{spacing}
\end{document}